\documentclass[10pt]{wlscirep}
\usepackage[resetlabels]{multibib}
\usepackage{amsfonts,amssymb,stmaryrd,latexsym,amsmath,braket}
\usepackage{graphicx,subfigure}
\usepackage{times}
\usepackage{slashed}
\usepackage{hyperref}
\newcites{X}{References}
\newcites{Y}{References}
\usepackage{comment}

\title{Emergent geometry and duality in the carbon nucleus}

\author[1]{Shihang~Shen}
\author[1,2]{Timo~A.~L{\"a}hde}
\author[3]{Dean~Lee} 
\author[4,1,2,5]{Ulf-G.~Mei{\ss}ner}
   
\affil[1]{Institut~f\"{u}r~Kernphysik,~Institute~for~Advanced~Simulation,~
J\"{u}lich~Center~for~Hadron~Physics,\protect\linebreak Forschungszentrum~J\"{u}lich,
~D-52425~J\"{u}lich,~Germany}

\affil[2]{Center for Advanced Simulation and Analytics (CASA),~Forschungszentrum~J\"{u}lich,~D-52425~J\"{u}lich,~Germany}

\affil[3]{Facility~for~Rare~Isotope~Beams~and~Department~of~Physics~and~Astronomy,~Michigan~State~University, East~Lansing~MI~48824,~USA}

\affil[4]{Helmholtz-Institut~f\"{u}r~Strahlen-~und~Kernphysik~and~Bethe~Center~for~Theoretical~Physics,\protect\linebreak
~Universit\"{a}t~Bonn,~D-53115~Bonn,~Germany}

\affil[5]{Tbilisi State University, 0186 Tbilisi, Georgia}


\begin{abstract}
The carbon atom provides the backbone for the complex organic chemistry composing the building blocks of life. The physics of the carbon nucleus in its predominant isotope, $^{12}$C, is similarly full of multifaceted complexity.  Some nuclear states of $^{12}$C can be preferentially treated as a collection of independent particles held by the mean field of the nucleus, while other states behave more as a collection of three alpha-particle clusters. But these two pictures are not mutually exclusive, and some states can be described in either fashion \cite{Kanada-Enyo:2006rjf,Kanada-Enyo:2018fjk}.  In this work, we provide the first model-independent tomographic scan of the three-dimensional geometry of the nuclear states of $^{12}$C using the {\it ab initio} framework of nuclear lattice effective field theory.
We find that the well-known but enigmatic Hoyle state is composed of a ``bent-arm'' or obtuse triangular arrangement of alpha clusters. We identify all of the low-lying nuclear states of $^{12}$C as having an intrinsic shape composed of three alpha clusters forming either an equilateral triangle or an obtuse triangle. From these basic structural formations, the various nuclear states correspond to different rotational and vibrational excitations as well as either distortions or large-amplitude displacements of the alpha clusters. The states with the equilateral triangle formation also have a dual description in terms of particle-hole excitations in the mean-field picture.  We compare our theoretical calculations with experimental data for binding energies, quadrupole moments, electromagnetic transitions, charge densities, and form factors.  The overall agreement is good, and further studies using higher-fidelity interactions are planned.
\end{abstract}

\begin{document}

\maketitle

{\em Introduction:}~The physics of the $^{12}$C nucleus is a fascinating subject with a long and fabled history \cite{Hoyle:1954zz,Cook:1957zz}, and recent groundbreaking experimental results have provided hints of new states with exotic structures \cite{Freer:2007zz,Freer:2009zz,Chernykh:2010zu,Itoh:2011zz,Freer:2011zza,Zimmerman:2013cxa,Marin-Lambarri:2014zxa,Smith:2020nkz,Bishop:2020sqi,Kuhlwein:2021zks,Li:2022syy}.  However, the underlying structures of several nuclear states of $^{12}$C remain without a consensus of agreement, and answers to such questions would provide deep insights into the emergent correlations relevant to nuclear binding and the panoply of possible structures that may appear in other nuclear systems.  The most famous example is the case of the so-called Hoyle state, and its hypothetical rotational band partners. The Hoyle state is a narrow resonance, whose close proximity to the energy threshold for three alpha particles greatly enhances the reaction rate of the triple-alpha process, which is key to the production of carbon in evolved, helium-burning stars \cite{Freer:2014qoa,Freer:2017gip}.  Much progress
has been made in understanding the spectrum of $^{12}$C including the Hoyle state, in theoretical studies using the no-core shell model \cite{Navratil:2007we,LENPIC:2018ewt}, symmetry-adapted no-core shell model \cite{Dreyfuss:2012us},
shell model \cite{Yuan:2012zz}, quantum Monte Carlo simulations (QMC) \cite{Carlson:2014vla},
replica exchange MC (RXMC) \cite{Ichikawa:2021zup}, 
antisymmetrized molecular dynamics (AMD) \cite{Kanada-Enyo:2006rjf,Kanada-Enyo:2015vwc}, fermion molecular dynamics (FMD) \cite{Chernykh:2007zz},
density functional theory \cite{Ebran:2014pda,Zhao:2014vfa,Ren:2018mbi},  Bose-Einstein condensate (BEC) wave functions \cite{Tohsaki:2001an,Funaki:2003af,Funaki:2004xe}, alpha cluster models (ACM) \cite{Chernykh:2007zz}, and nuclear lattice effective field theory (NLEFT)  \cite{Epelbaum:2011md,Epelbaum:2012qn}.  There are two main impediments to reaching definitive conclusions about the structure of the low-lying $^{12}$C states. The first is the inability to perform calculations that can handle strong multi-particle correlations. The second is the inability to measure the detailed spatial correlations required to determine the intrinsic structure of the twelve-particle wave function.  In this work we address both problems.  We perform unconstrained lattice Monte Carlo simulations using the framework of NLEFT \cite{Lee:2008fa,Lahde:2019npb}, including all possible multi-particle quantum correlations. From these simulations, we determine the full twelve-particle correlations and use a model-independent tomographic projection to determine the intrinsic three-dimensional structure of each nuclear state.

In our calculations, we use a simple interaction between the nucleons that is independent of spin and whether the nucleon is a proton or neutron, \textit{i.e.}, of isospin. Conveniently, this is similar to the interactions used in most AMD and BEC calculations.  Therefore, any significant differences in the conclusions are likely due to differences in the computational method, or in the choice of calculated observables, rather than to differences in the nucleon-nucleon interaction. The same type of lattice interaction has been used to describe the ground state energies of light and medium-mass nuclei~\cite{Lu:2018bat} and the thermodynamics of symmetric nuclear matter~\cite{Lu:2019nbg}. Perhaps most importantly, it reproduces the low-energy spectrum of $^{12}$C well~\cite{Shen:2021kqr}. 

{\em Lattice methods:} For the lattice simulations presented here, we use a spatial lattice spacing $a = 1.64$~fm and a temporal lattice spacing of $a_t = 0.55$~fm$/c$, where $c$ is the speed of light.  While the individual nucleons must reside on lattice sites, the center of mass (c.m.) of the $^{12}$C nucleus is constrained to a much finer three-dimensional grid of lattice spacing $a/12 =0.137$~fm, which equals the resolution of our tomographic projection for each $^{12}$C state.
The lattice interaction has the form
\begin{equation}\label{eq:vsu4}
  V = \frac{C_2}{2!} \sum_{\mathbf{n}}
  \tilde{\rho}(\mathbf{n})^2 + \frac{C_3}{3!} \sum_{\mathbf{n}}
  \tilde{\rho}(\mathbf{n})^3,
\end{equation}
where $C_2$ and $C_3$ are the two-body and the three-body interaction coefficients, respectively.
The vector $\mathbf{n}$ denotes the spatial lattice sites. The definition of the smeared density operator
$\tilde{\rho}(\mathbf{n})$ is given in Methods, and it entails two parameters, $s_{\rm L}$, and $s_{\rm NL}$.  The four parameters $C_2, C_3, s_{\rm L}$, and $s_{\rm NL}$ are determined by a joint fit to the ground-state energies of $^{4}$He and $^{12}$C, to the ground-state charge radius of $^{12}$C, and to several electromagnetic transition rates.

A summary of the auxiliary-field Monte Carlo calculations is given in Methods.  We use an assortment of different initial states for each state of $^{12}$C and verify that our choice of initial state does not affect the final observables. The initial states we consider include a wide variety of mean-field states composed of products of harmonic oscillator (shell model) states as well as different geometric configurations of alpha clusters. We use the pinhole algorithm to determine the probability distribution for the nucleon positions, spins, and isospins \cite{Elhatisari:2017eno}.  For each pinhole configuration, we know the positions of all $A$ nucleons, and thus the position of each nucleon relative to the center of mass is easily calculated.  From this information, we can compute any observable that does not involve displacements of the nucleons.  For example, the electric charge density can be determined from the distribution of protons, taking into account the charge radius of protons. The charge form factor $F(q)$ is then calculated by performing the Fourier transform of the charge density.

\begin{figure}[!tbp]
\centering
\raisebox{0.05\height}
{\includegraphics[width=0.64\columnwidth]{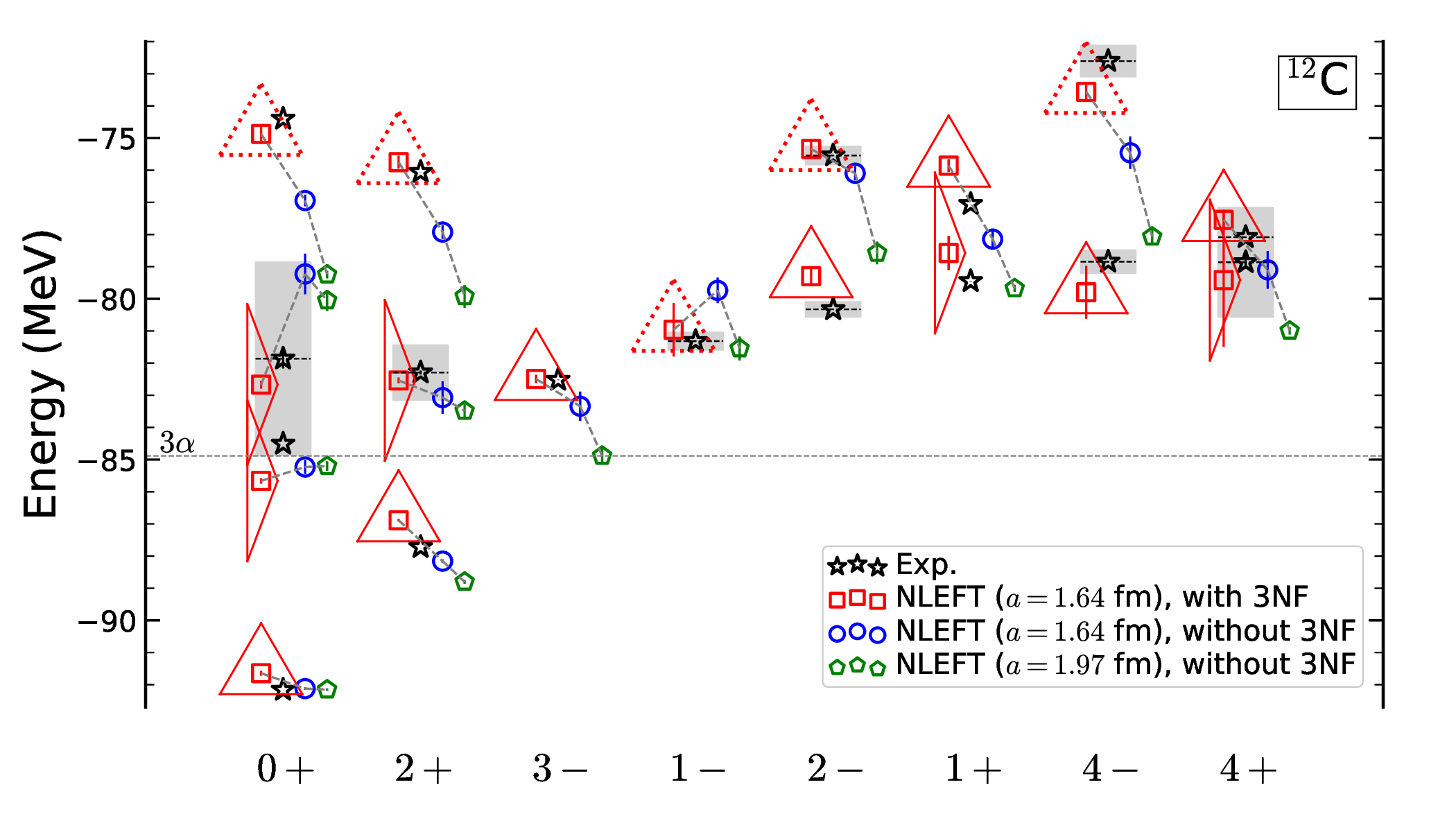}}
\includegraphics[width=0.35\columnwidth]{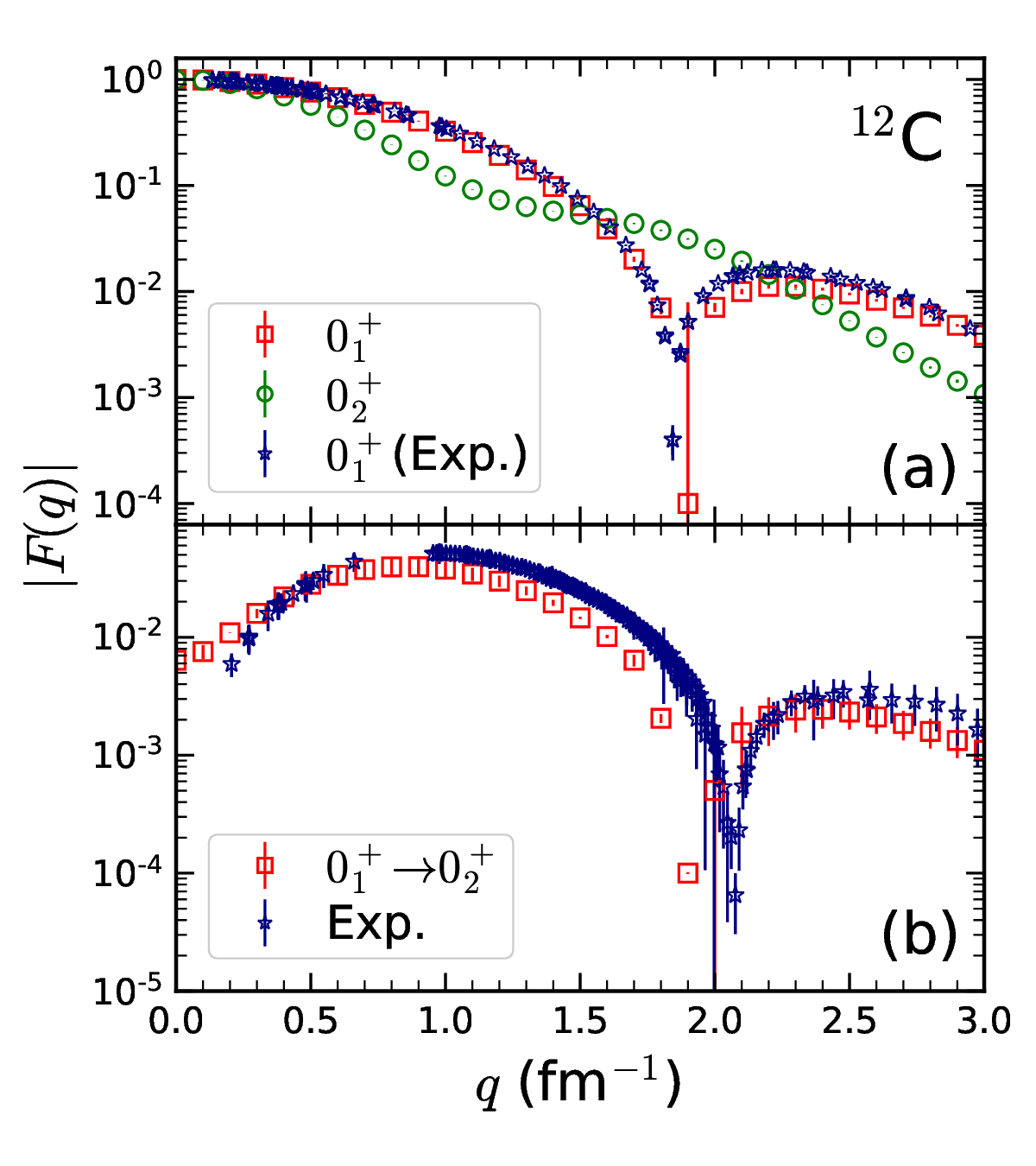}
\caption{\label{fig:spectrum_form} {\bf Left Panel:} Spectrum of $^{12}$C (red squares) in comparison with experimental data (black stars).  The error bars correspond to one standard deviation errors.  The grey shaded regions indicate decay widths for cases where it has been measured.  Previous results for lattice spacing $a = 1.64$~fm (blue circles) and $a = 1.97$~fm (green pentagons) are also shown~\cite{Shen:2021kqr}.  The triangular shapes indicate the intrinsic shape of each nuclear state, either equilateral or obtuse triangle arrangements of alpha clusters. The dotted lines for some equilateral triangles indicate significant distortions or large-amplitude displacements of the alpha clusters. {\bf Right Panel:} The absolute value of the charge form factor $F(q)$.
The top figure (a) shows the ground state (red squares) and Hoyle state (green circles), and the bottom figure (b) shows the transition from the ground state to the Hoyle state (red squares). The error bars correspond to one standard deviation errors.  Experimental data (purple stars) are shown for comparison~\cite{Sick:1970ma,Strehl:1970,Crannell:2005wer,Chernykh:2010zu}. }
\end{figure}

{\em Spectrum and electromagnetic properties:}~We have calculated the $^{12}$C spectrum up to excitation energies of about $15$~MeV.  The results are plotted as red squares in the left panel of Fig.~\ref{fig:spectrum_form} for different values of the angular momentum and parity.
For comparison we show the experimental data (black stars) \cite{Kelley:2017qgh} and results from our previous work with lattice spacing $a = 1.64$~fm (blue circles) and $a = 1.97$~fm (green pentagons), which were performed without three-nucleon forces \cite{Shen:2021kqr}.
Overall, the agreement with the empirical results is quite good. The triangular shapes surrounding the data points are explained later in our discussion. In the right panel of Fig.~\ref{fig:spectrum_form}, we show the form factors for the ground state and the Hoyle state in the top figure (a), and the transition form factor from the ground state to the Hoyle state in the bottom figure (b). For comparison, we show the experimental data for the ground state and transition form factors \cite{Sick:1970ma,Strehl:1970,Crannell:2005wer,Chernykh:2010zu}, and the agreement is fairly good.

\begin{table}[!tp]
    \centering
    \caption{Energies and charge radii $r_c$ (or point radii $r$) of $^{12}$C calculated by NLEFT,
      compared to  calculations from FMD \cite{Chernykh:2007zz}, ACM
      \cite{Chernykh:2007zz}, BEC\cite{Funaki:2003af,Funaki:2004xe}, and RXMC \cite{Ichikawa:2021zup} as well as
      experiment \cite{Kelley:2017qgh,Angeli:2013epw}. All energies are in MeV and radii in fm.
      For $r_c(0_1^+)$, the charge radius of the proton  $r_E^p = 0.84$~fm~\cite{Lin:2021xrc} is 
      added in quadrature. For the NLEFT results, 
      the first error bars are one standard deviation estimates due to stochastic errors and Euclidean time extrapolation. The second error bars are an estimate of systematic errors due to broken rotational invariance from the finite periodic volume and nonzero lattice spacing.  Additional systematic errors due to the choice of interaction are described in Methods.}
    \begin{tabular}{|l|ccccc|c|}
    \hline                                                               
                 & NLEFT  & FMD    & $\alpha$ cluster & BEC & RXMC & Exp.  \\
    \hline                                                                   
    $E(0_1^+)$   & $-91.6(1)$ & $-92.6$ & $-89.6$     & $-89.5$ & $-88.0$ & $-92.2$ \\
    $E(0_2^+)$   & $-85.7(1)$ & $-83.1$ & $-81.7$     & $-81.8$ & $-81.4$ &  $-84.5$ \\
    $E(0_3^+)$   & $-82.7(1)$ & $-80.7$ & $-79.2$     & -- & $-79.0$ &  $-81.9(3)$ \\
    $E(2_1^+)$   & $-86.9(1)(1.5)$ & $-87.3$ & $-87.0$     & $-86.7$ &  -- & $-87.7$      \\
    $E(2_2^+)$   & $-82.5(1)(2.1)$ & $-80.8$ & $-80.4$   & $-80.5$ & -- & $-82.3$  \\
    $r_c(0_1^+)$ & $2.54(1)$   & $2.53$   & $2.54$             & $2.53$ & $2.65$ & $2.47(2)$  \\
    $r(0_2^+)$   & $3.45(2)$   & $3.38$   & $3.71$             & $3.83$ & $4.00$  & -- \\
    $r(0_3^+)$   & $3.47(1)$   & $4.62$   & $4.75$             & -- & $4.80$  & -- \\
    $r(2_1^+)$   & $2.42(1)(1)$   & $2.50$   & $2.37$             & $2.38$ &  -- & --    \\
    $r(2_2^+)$   & $3.30(1)(4)$   & $4.43$   & $4.43$             & -- & -- & --      \\
    \hline
    \end{tabular}
    \label{tab:energy}
\end{table}

In Table~\ref{tab:energy}, the energies and radii of
$0_1^+, 0_2^+, 0_3^+, 2_1^+$, and $2_2^+$ states are shown, in comparison with other theoretical calculations \cite{Chernykh:2007zz,Funaki:2003af,Funaki:2004xe,Ichikawa:2021zup}
and experimental data~\cite{Kelley:2017qgh,Angeli:2013epw}. Our $J^\pi_n$ notation indicates the angular momentum $J$, parity $\pi$, and ordinal number $n$.
The NLEFT energies and radii agree very well with empirical results.  Since the $0_2^+, 0_3^+$ and $2_2^+$ states are unbound or nearly unbound with respect to the three-alpha threshold, the radii are calculated when the corresponding state is placed in a periodic cube with length $L = 14.8$~fm.  The first error bars are one standard deviation estimates due to stochastic errors and Euclidean time extrapolation uncertainties.  The second error bars are an estimate of systematic errors due to broken rotational invariance from the finite periodic volume and nonzero lattice spacing.  Additional systematic errors due to the choice of interaction are described in Methods.  In Table~\ref{tab:trans}, the electric quadrupole moments of the $2^+$ states and electric transition rates involving the $0_1^+, 0_2^+, 0_3^+,$ and $2_1^+$ states of $^{12}$C
obtained by NLEFT are given in comparison with other theoretical calculations \cite{Chernykh:2007zz,DAlessio:2020aek} and experimental data \cite{Ajzenberg-Selove:1990fsm,SaizLomas:2021xzn}.  While the quadrupole transitions have significant errors due to broken rotational invariance, the overall agreement with empirical results is good.
\begin{table}[!tb]
    \centering
    \caption{Quadrupole moment and transition rates of $^{12}$C calculated by NLEFT, compared to calculations based on
      FMD \cite{Chernykh:2007zz}, $\alpha$ cluster models  \cite{Chernykh:2007zz}, in-medium no-core shell model (NCSM) \cite{DAlessio:2020aek}, generator coordinate method (GCM) calculation~\cite{Zhou:2016mhb}
      and experiment \cite{Ajzenberg-Selove:1990fsm,SaizLomas:2021xzn}.
      Units for $Q$ and $M(E0)$ are $e\,$fm$^2$,
      and for $B(E2)$ $e^2\,$fm$^4$.
      For the NLEFT results, the first errors bar refer to the Euclidean time extrapolation uncertainties.  The second error bars are an estimate of errors due to broken rotational invariance from the finite periodic volume and nonzero lattice spacing. Additional systematic errors due to the choice of interaction are described in Methods.
      }
    \begin{tabular}{|l|ccccc|c|}
    \hline
    & NLEFT & FMD & $\alpha$ cluster & NCSM & GCM & Exp. \\
    \hline
    $Q(2_1^+)$ & $6.8(3)(1.2)$ & -- & -- & $6.3(3)$ & -- &  $8.1(2.3)$  \\
    $Q(2_2^+)$ & $-35(1)(1)$ & -- & -- & -- & -- & -- \\
    $M(E0,0_1^+\to 0_2^+)$ & $4.8(3)$  & $6.5$ & $6.5$ & -- & $6.2$ & $5.4(2)$ \\
    $M(E0,0_1^+\to 0_3^+)$ & $0.4(3)$  & -- & -- & -- & $3.6$ & -- \\
    $M(E0,0_2^+\to 0_3^+)$ & $7.4(4)$  & -- & -- & -- & $ 47.0$ & -- \\
    $B(E2,2_1^+\to 0_1^+)$ & $11.4(1)(4.3)$  & $8.7$ & $9.2$ & $8.7(9)$ & -- & $7.9(4)$\\
    $B(E2,2_1^+\to 0_2^+)$ & $2.4(2)(7)$  & $3.8$ & $0.8$ & -- & -- & $2.6(4)$\\
    \hline
    \end{tabular}
    \label{tab:trans}
\end{table}

{\em Intrinsic structures:}~We compute the intrinsic structure of the nuclear states in the following manner.  We first select pinhole configurations which contain exactly three spin-up protons. For the spin-independent interactions used in this work, this corresponds to all configurations. For each spin-up proton, we locate the closest spin-down proton, spin-up neutron, and spin-down neutron. We identify these four nucleons as an alpha cluster, and determine its center of mass. In this manner, we locate the positions of three alpha clusters for each nuclear state of $^{12}$C.  The root-mean-square (RMS) matter radii of the alpha clusters defined in this manner range from $1.57$~fm to $1.62$~fm for the low-lying states in $^{12}$C.  Since this is very close to the matter radius of an isolated alpha particle using the same interactions, $r_\alpha =1.63$ fm, we conclude that our process of identifying alpha cluster configurations is accurate and free of significant artifacts.

\begin{figure}[!htbp]
\centering
\includegraphics[width=0.49\columnwidth]{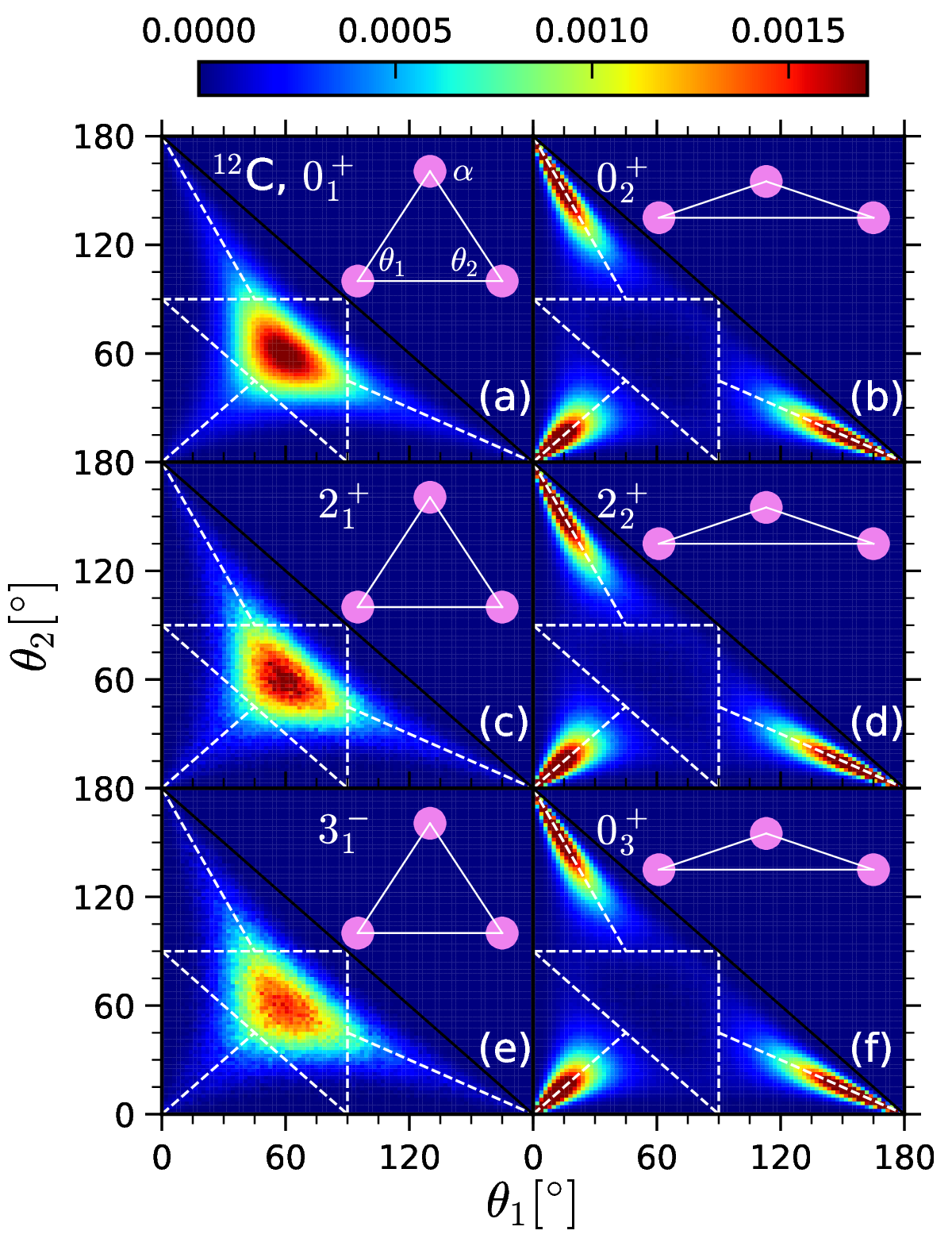}
\includegraphics[width=0.49\columnwidth]{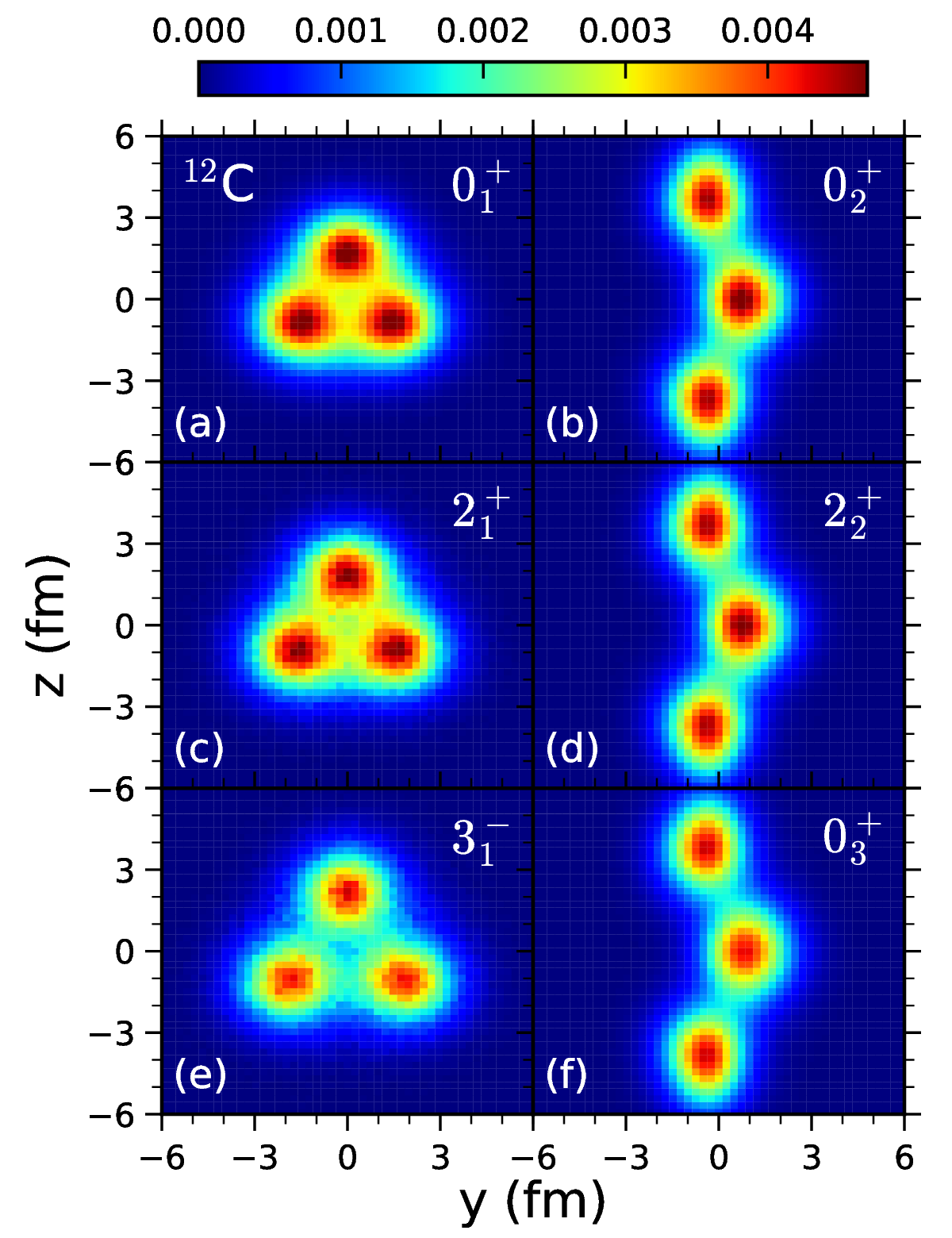}\caption{\label{fig:theta12_dstx0} Nuclear density distributions for the (a) $0_1^+$ ground state, (b) $0_2^+$ Hoyle state, (c) $2_1^+$, (d) $2_2^+$, (e) $3_1^-$, and (f) $0_3^+$ states. The red (blue) color signals
a high (low) probability. {\bf Left Panel:} Density distribution for the two inner angles of the triangle formed by the three alpha clusters. The two axes are for the two inner angles $\theta_1$ and $\theta_2$ measured in degrees.
{\bf Right Panel:} Tomographic projection of the nuclear density for different states of $^{12}$C.  In each case the orientation of the shortest root-mean-square direction is aligned with the $x$ axis.}
\end{figure}

The three alpha clusters that we have identified define a triangle in three-dimensional space with interior angles $\theta_1$, $\theta_2$, and $180^\circ - \theta_1-\theta_2$.  In the left panel of Fig.~\ref{fig:theta12_dstx0}, we show the probability distributions as a function of $\theta_1$ and $\theta_2$ for (a) the $0_1^+$ ground state, (b) $0_2^+$ Hoyle state, (c) $2_1^+$, (d) $2_2^+$, (e) $3_1^-$, and (f) $0_3^+$ states.  The black solid line at $\theta_2 =
180^\circ - \theta_1$ separates the physical region (lower left) and the unphysical region (upper right).  The dashed white triangle formed by the line segments $\theta_1=90^\circ$, $\theta_2=90^\circ$, and $\theta_2=90^\circ-\theta_1$, represents cluster configurations that are right triangles. The interior region of the dashed white triangle corresponds to configurations that are acute triangles, and the exterior region corresponds to obtuse triangles.  The other three white dashed line segments along the lines $\theta_1=\theta_2$, $\theta_1=\theta_3$, and $\theta_2=\theta_3$ represent cluster configurations that are obtuse isosceles triangles. For the $0^+_1$ ground state, the probability distribution is strongly centered around an equilateral triangle, $\theta_1=\theta_2=\theta_3 = 60^\circ$. The $2_1^+$ and $3_1^-$ states have similar equilateral triangular shapes. In contrast, the $0^+_2$ Hoyle state corresponds to an obtuse isosceles triangle. This finding is consistent with older NLEFT studies \cite{Epelbaum:2011md,Epelbaum:2012qn}. The $2_2^+$ and $0^+_3$ states also have obtuse isosceles triangular shapes.

We now define a model-independent tomographic projection of the three-dimensional nuclear density for the states of $^{12}$C. In order to construct this projection, we first identify the $x$ axis as the direction with the smallest RMS deviation of the nucleon positions relative to the center of mass.
For the nuclear states that we have already identified as having an equilateral triangular shape, we rotate the density configurations along the $x$ axis so that one of the three clusters is pointing along the positive $z$ direction.  We then symmetrize with respect to $0^\circ$, $120^\circ$ and $240^\circ$ rotations about the $x$ axis. For nuclear states that we have already identified as having an obtuse isosceles shape, we identify the $z$ axis as the direction with the longest RMS deviation of the nucleon positions relative to the c.m.  We then rotate the density configurations along the $z$ axis so that the alpha cluster with the smallest $z$ value has a positive $y$ coordinate.

In the right panel of Fig.~\ref{fig:theta12_dstx0}, we show the density distribution of selected states of $^{12}$C prepared in the model-independent manner described above. The $0_1^+$, $2_1^+$, $3_1^-$, $4_1^-$, and $4_2^+$ states (see Methods for the last two) have similar intrinsic equilateral triangular shapes, consistent with an interpretation as members of a rotational band built on top of the $0_1^+$ state. The $0_2^+$, $2_2^+$, $4_1^+$ states (see Methods for the last one) have similar intrinsic obtuse isosceles triangle shapes and are consistent with belonging to a rotational band built on top of the $0_2^+$ state. These findings are consistent with previous studies in the literature based on group theoretical considerations \cite{Marin-Lambarri:2014zxa}. We note that models where the Hoyle state has an equilateral triangle symmetry predict an additional $3^-$ and $4^-$ state in the Hoyle state rotational band.

The $0_3^+$ state has been discussed as a breathing mode excitation of the Hoyle state \cite{Li:2020zsd,Li:2022syy,Zhou:2016mhb}, but its detailed structure remains a matter of debate.  For example, in a recent work~\cite{Li:2020zsd,Li:2022syy} the $0_3^+$ and Hoyle states are suggested to have an equilateral triangular shape. A gas-like structure with a very large radius has also been predicted~\cite{Zhou:2016mhb} for the $0_3^+$ state. Our lattice findings suggest that the $0_3^+$ state is a small-amplitude vibrational excitation of the Hoyle state. Our findings for the intrinsic shapes of the low-lying states of $^{12}$C are summarized by the triangular shapes in the left panel of Fig.~\ref{fig:spectrum_form}. The triangular symbols indicate the intrinsic shape of each nuclear state, either equilateral or obtuse triangle arrangements of alpha clusters. The dotted lines for some equilateral triangles indicate significant distortions or large-amplitude displacements of the alpha clusters. Examples of these states are shown in Methods.  We find that all of the low-lying states with an equilateral triangle formation have significant overlap with some initial state composed of an antisymmetrized product of mean-field shell model states.  This constitutes the aforementioned duality between shell model and cluster states. In contrast, all of the low-lying states with an obtuse isosceles triangle formation have very little overlap with shell model initial states. 

In summary, we have presented the first model-independent tomographic scan of the three-dimensional geometry of the nuclear states of $^{12}$C using the {\it ab initio} framework of nuclear lattice effective field theory.  We find that the Hoyle state and its $2^+_2$ and $4^+_1$ rotational excitations are composed of an obtuse isosceles triangular arrangement of alpha clusters.  All of the low-lying nuclear states of $^{12}$C have an intrinsic shape composed of three alpha clusters forming either an equilateral triangle or an obtuse triangle. From these basic structural formations, the various nuclear states correspond to different rotational and vibrational excitations as well as either distortions or large-amplitude displacements of the alpha clusters. Future studies are planned to revisit this analysis using high-fidelity chiral effective field theory interactions.

{\it We are grateful for discussions with members of the Nuclear Lattice Effective Field Theory Collaboration as well as Scott Bogner, Jerry Draayer, Martin Freer, Heiko Hergert, Morten Hjorth-Jensen, and Witek Nazarewicz.
We acknowledge funding by  the Deutsche Forschungsgemeinschaft
(DFG, German Research Foundation) and the NSFC through the funds provided  to  the  Sino-German
Collaborative  Research  Center  TRR110  ``Symmetries  and  the  Emergence  of  Structure in  QCD''
(DFG  Project  ID 196253076  -  TRR  110,  NSFC Grant  No.  12070131001),
the Chinese Academy of Sciences (CAS) President's International Fellowship Initiative (PIFI)
(Grant No. 2018DM0034), Volkswagen Stiftung  (Grant  No.  93562),  the European Research Council (ERC) under the
European Union's Horizon 2020 research and innovation programme (ERC AdG EXOTIC, grant agreement No. 101018170)
and the U.S. Department of Energy (DE-SC0013365 and DE-SC0021152) and the Nuclear Computational Low-Energy
Initiative (NUCLEI) SciDAC-4 project (DE-SC0018083) as well as computational resources provided by the Gauss
Centre for Supercomputing e.V. (www.gauss-centre.eu) for computing time on the GCS Supercomputer JUWELS at J{\"u}lich
Supercomputing Centre (JSC) and special GPU time allocated on JURECA-DC as well as
the Oak Ridge Leadership Computing Facility through the INCITE award
``Ab-initio nuclear structure and nuclear reactions''.}


\bibliographystyle{apsrev}

\noindent {\bf \large Author Contributions} S.~S. performed the theoretical work, adapted codes developed by the NLEFT collaboration, and performed the required lattice MC simulations. T.~L., D.~L., and U.-G.~M. supervised the direction of the research. All authors were involved in the writing, editing, and review of this work.
\\
\\
\noindent {\bf \large Author Information}  The authors declare no competing financial interests.

\newpage 

\section*{Methods}
\renewcommand{\thefigure}{S\arabic{figure}}
\renewcommand{\theequation}{S\arabic{equation}}
\setcounter{figure}{0}
\setcounter{equation}{0}

\subsection*{Auxiliary-field lattice Monte Carlo}

We start from a set of trial initial states $|\Phi_i\rangle$ with $i = 1, 2, \dots, N_{\rm ch}$, for $N_{\rm ch}$ channels, and perform Euclidean time projection with the transfer matrix $M = :\exp(-\alpha_t H):$, where the colons denote normal ordering and 
$\alpha_t = a_t/a$. The Hamiltonian is $H=T+V$, where the
kinetic energy $T$ is $\mathbf{p}^2/(2m_N)$ and the interaction $V$ is defined in the main text in Eq.~\eqref{eq:vsu4}. The Euclidean projection amplitude at time step $N_t$
is $Z_{kl}(N_t) = \langle \Phi_k| M^{N_t} |\Phi_l \rangle$.
Each of the trial states $|\Phi_i\rangle$ is a Slater determinant of single-particle wave functions.
For the cluster initial states, we consider an antisymmetrized product of alpha clusters with spatially distributed Gaussian wave packets,
$\phi(\mathbf{r}) = \exp\left(-{\mathbf{r}^2}/{2w^2}\right)$, with $w = 1.6 - 1.8\,$fm for the width of the wave packet.
Each wave function is projected onto specific
irreducible representations ({\em irrep}) of the cubic group \cite{Johnson:1982yq,Lu:2014xfa}.
More details on such trial states can be found in Ref.~\cite{Shen:2021kqr_S}.
The energies at Euclidean time step $N_t$ are obtained from the eigenvalues of the adiabatic transfer matrix,
\begin{equation}
M_{qq'}^{(a)}(N_t) = \sum_{q''} Z_{qq''}^{-1}(N_t) Z_{q''q'}^{}(N_t+1). 
\end{equation}
We convert the eigenvalues $\lambda_i(N_t)$ to energies using the relation
$\exp(-\alpha_t E_i(N_t)) = \lambda_i(N_t)$.
In the calculation, the two-body interactions among $A$ nucleons are replaced by one-body interactions
between single nucleons with auxiliary fields that are sampled by Monte Carlo methods \cite{Lahde:2019npb_S}. 

\subsection*{Pinhole algorithm}

We use the pinhole algorithm (PA)~\cite{Elhatisari:2017eno} to determine a classical distribution for the nucleon positions, spins, and isospins. For this, we compute the amplitude
\begin{equation}
\label{eq:Zkl2}
Z_{kl}(\mathbf{n}_1,\dots,\mathbf{n}_A,N_t) = \langle \Phi_k| M^{N_t/2} \rho(\mathbf{n}_1,\dots,
\mathbf{n}_A)M^{N_t/2} |\Phi_l \rangle,
\end{equation}
with the $A$-body density operator $\rho(\mathbf{n}_1,\dots,\mathbf{n}_A)$ constructed from the normal-ordered product of density operators $ \rho(\mathbf{n}_i) = a_{i}^\dagger(\mathbf{n}) a_i^{}(\mathbf{n})$.
Since the SU(4)-symmetric interaction used here does not change spin or isospin, such indices have been suppressed. The positions for $A$ nucleons (or ``pinholes'') $\mathbf{n}_i$ are sampled stochastically. The finite size of the nucleons is
accounted for by a random Gaussian smearing of the
nucleon positions.

\subsection*{Nucleon density operators}

The definition of the density operator $\tilde{\rho}(\mathbf{n})$ that appears in main text as Eq.\eqref{eq:vsu4} is given by \cite{Elhatisari:2017eno_S}:
\begin{equation}
  \tilde{\rho}(\mathbf{n}) = \sum_{i=1}^A \tilde{a}_i^\dagger(\mathbf{n}) \tilde{a}_i^{}(\mathbf{n})
  + s_L \sum_{|\mathbf{n}'-\mathbf{n}|=1}\sum_{i=1}^A \tilde{a}_i^\dagger(\mathbf{n}')\tilde{a}_i^{}(\mathbf{n}').
\end{equation}
Here $s_L$ is the local smearing parameter, and the smeared annihilation operator is
\begin{equation}
  \tilde{a}_{i}(\mathbf{n}) = a_i(\mathbf{n}) + s_{\rm NL} \sum_{|\mathbf{n}'-\mathbf{n}|=1} a_i(\mathbf{n}'),
\end{equation}
where $s_{\rm NL}$ is the non-local smearing parameter.

\subsection*{Euclidean time extrapolation}

\begin{figure}[t]
\centering
\includegraphics[width=0.5\columnwidth]{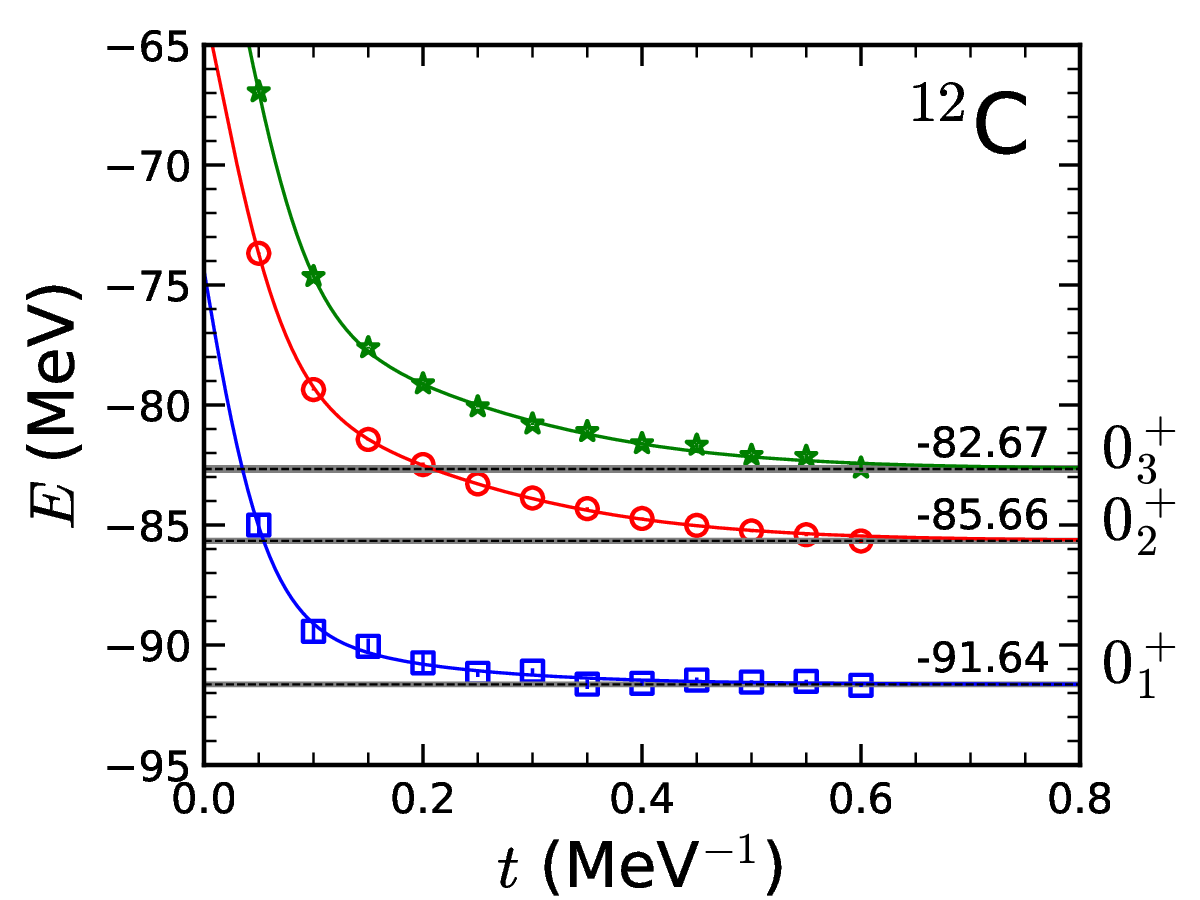}
\caption{\label{fig:econv} Transient energies (open symbols) of the lowest three 0+ states, as a function
of Euclidean projection time. The example shown is a 
three-channel auxiliary field lattice MC calculation using alpha-cluster trial states and $A_1^+$ projection. The data are fitted and extrapolated using Eq.~\eqref{eq:extrap}. The widths of the horizontal lines reflect the extrapolation error.}
\end{figure}

In NLEFT, energies and operator expectation values are computed using auxiliary-field lattice MC at finite Euclidean projection time. As an example, the low-energy spectrum is obtained from the transient energies $E_i(N_t) = -{\log(\lambda_i(N_t))}/{\alpha_t}$. Some extrapolation is then usually required to obtain the corresponding values at infinite projection time. For instance, we obtain energies at infinite projection time by fitting the lattice data using the \textit{ansatz}~\cite{Lahde:2014sla_S,He:2019ipt}
\begin{equation}
\label{eq:extrap}
  E_i(t) = \frac{E_i+\displaystyle\sum_{k=1}^{k_{\rm max}} (E_i+\Delta E_{i,k}) c_{i,k}e^{-\Delta E_{i,k} t}}
  {1+\displaystyle\sum_{k=1}^{k_{\rm max}} c_{i,k}e^{-\Delta E_{i,k} t}},
\end{equation}
where $E_i, \Delta E_{i,k}, c_{i,k}$ are fit parameters. The number of exponentials $k_{\rm max}$ should be kept as small as possible~\cite{Lahde:2014sla_S}, in order to avoid
fitting statistical fluctuations.
For expectation value of operators of the type $\langle i|\hat{O}|i \rangle$, or transition operators
$\langle i|\hat{O}|j \rangle$, we use the lowest order extrapolation formulas~\cite{He:2019ipt}
\begin{equation}
  O_i(t) = \frac{O_i+O_{i,1}e^{-\Delta E_{i,1} t/2}+O_{i,2}e^{-\Delta E_{i,1} t}}{1+c_{i,1}e^{\Delta E_{i,1}t}}, \quad\quad
  O_{ij}(t) = \frac{O_i+O_{i,1}e^{-\Delta E_{i,1} t/2}+O_{j,1}e^{-\Delta E_{j,1} t/2}}{1+c_{i,1}e^{\Delta E_{i,1}t}+c_{j,1}e^{\Delta E_{j,1}t}},
\end{equation}
with only one $\Delta E_{i,k}$ (or $\Delta E_{j,k}$).
In Fig.~\ref{fig:econv}, we show the extrapolation using Eq.~(\ref{eq:extrap}) for the lowest three $0^+$ states of $^{12}$C. With $k_{\rm max} = 2$, the convergence of the auxiliary-field MC data is well described.

\subsection*{Interaction parameters and systematic errors}

In Table~\ref{tab:fit-v3}, we summarize the interaction parameters, together with the observables used to determine them. Our preferred set of parameters is denoted V1. It is then of interest to assess the systematic error associated with this choice of parameters. Since we are here considering an SU(4) symmetric interaction without a specific EFT power counting scheme, our strategy to estimate the systematic errors is to perform
a number of fits with a different choice of the non-local smearing parameter $s_{\rm NL}$. The remaining three parameters $C_2,C_3,s_{\rm L}$ are again, in each case, fitted to the ground state energies of $^4$He and $^{12}$C, and to the electromagnetic radius of the $^{12}$C ground state. While this procedure allows for a large number of possibilities, we have chosen two interactions denoted V2 and V3, for closer comparison.

The results of our systematic error analysis are given in Table~\ref{tab:fit-v3}. For simplicity, calculated energies and matrix elements are given at finite Euclidean projection time, as the additional computational effort required for a full extrapolation would not substantially alter the results nor the conclusions. Based on earlier experience, we take $t = 0.4\,$MeV$^{-1}$ for $^{3}$H and $^{12}$C, and $t = 0.2\,$MeV$^{-1}$ for $^{4}$He. Such projection times should suffice for the purposes of the error analysis. 
\begin{table}[ht!]
  \caption{Summary of interaction parameters 
  $C_2, C_3, s_{\rm L}$, and $s_{\rm NL}$, along with an estimation of the systematic uncertainty due to the choice of interaction. The column labeled V1 denotes the preferred interaction used in this work, while V2 and V3 are used to quantify the systematic uncertainty. Observables shown are the ground state energies
    of $^{3}$H, $^{4}$He and $^{12}$C, the energy of the Hoyle state, and the charge radii of
    $^4$He and $^{12}$C calculated by $r_c = \sqrt{r^2+(0.84~\text{fm})^2}$. The interaction parameters are fitted to $E_{^{4}\text{He}}$,  $E_{^{12}\text{C}}$ and $r_{^{12}\text{C}}$. As a measure of the resulting systematic error, we give $Q(2_1^+)$ along with the $E0$ and $E2$ transition matrix elements, for each interaction. For these, results at finite Euclidean time are shown for simplicity. The errors in the round brackets are purely statistical.}
  \label{tab:fit-v3}
  \centering
  \begin{tabular}{|l|c|cc|}
  \hline
  & V1 & V2 & V3 \\
  \hline
  $s_{\rm NL}$  & $0.05$ & $0.1$   & $0.2$   \\
  $s_{\rm L}$   & $0.08$ & $0.071$ & $0.06$ \\
  $C_2$ [MeV$^{-2}$] & $-2.15 \times 10^{-5}$ & $-1.11 \times 10^{-5}$ & $-3.47 \times 10^{-6}$  \\
  $C_3$ [MeV$^{-5}$] & $6.17\times 10^{-12}$ & $-5.92 \times 10^{-13}$ & $-1.46 \times 10^{-12}$  \\
  \hline
  $E_{^{4}\text{He}}$ [MeV] & $-28.1 (1)$ & $-28.3 (1)$ & $-27.3 (1)$ \\
  $E_{^{12}\text{C}}$ [MeV] & $-91.6 (1)$ & $-91.8 (1)$ & $-90.7 (2)$ \\
  $r_{c,~^{12}\text{C}}$ [fm]  & $2.52 (1)$ & $2.55 (1)$  & $2.58 (1)$ \\
  \hline
  $E_{\text{Hoyle}}$ [MeV] & $-84.2 (1)$ & $-84.8 (5)$ & $-83.2 (11)$ \\
  $E_{^{3}\text{H}}$ [MeV] & $-10.1 (1)$ & $-8.1 (1)$ & $-5.1 (1)$ \\
  $r_{c,~^{4}\text{He}}$ [fm] & $1.63 (1)$ & $1.63 (1)$ & $1.64 (1)$ \\
  \hline
  $Q(2_1^+)$ [$e\,\mathrm{fm}^2$] & $6.9 (3)$ & $ 7.2 (4)$ & $ 6.1 (8)$ \\
  $M(E0,0_1^+\to 0_2^+)$ [$e\,\mathrm{fm}^2$] & $ 4.3 (3)$ & $ 2.9 (3)$ & $ 5.9 (7)$ \\
  $B(E2,2_1^+\to 0_1^+)$ [$e^2 \mathrm{fm}^4$] & $10.3 (2)$ & $10.7 (3)$ & $12.0 (5)$ \\
  $B(E2,2_1^+\to 0_2^+)$ [$e^2 \mathrm{fm}^4$] & $ 1.8 (1)$ & $ 3.6 (2)$ & $ 4.1 (5)$ \\
  \hline
  \end{tabular}
\end{table}
From Table~\ref{tab:fit-v3}, we find that all three interactions give a similarly good description of the fitted observables. While the energy of the Hoyle state and the $^4$He radius appear insensitive to the choice of interaction, the triton energy is influenced more. In particular, we observe that smaller values of $s_{\rm NL}$ give a more deeply bound $^3$H. We find that such systematic errors for the $^{12}$C spectrum do not exceed 2\%, while for the alpha particle they are about $2\ldots3\%$. The radii of $^4$He and $^{12}$C are even less affected ($\leq 1\%$).
\begin{figure}[ht!]
\centering
\includegraphics[width=0.5\columnwidth]{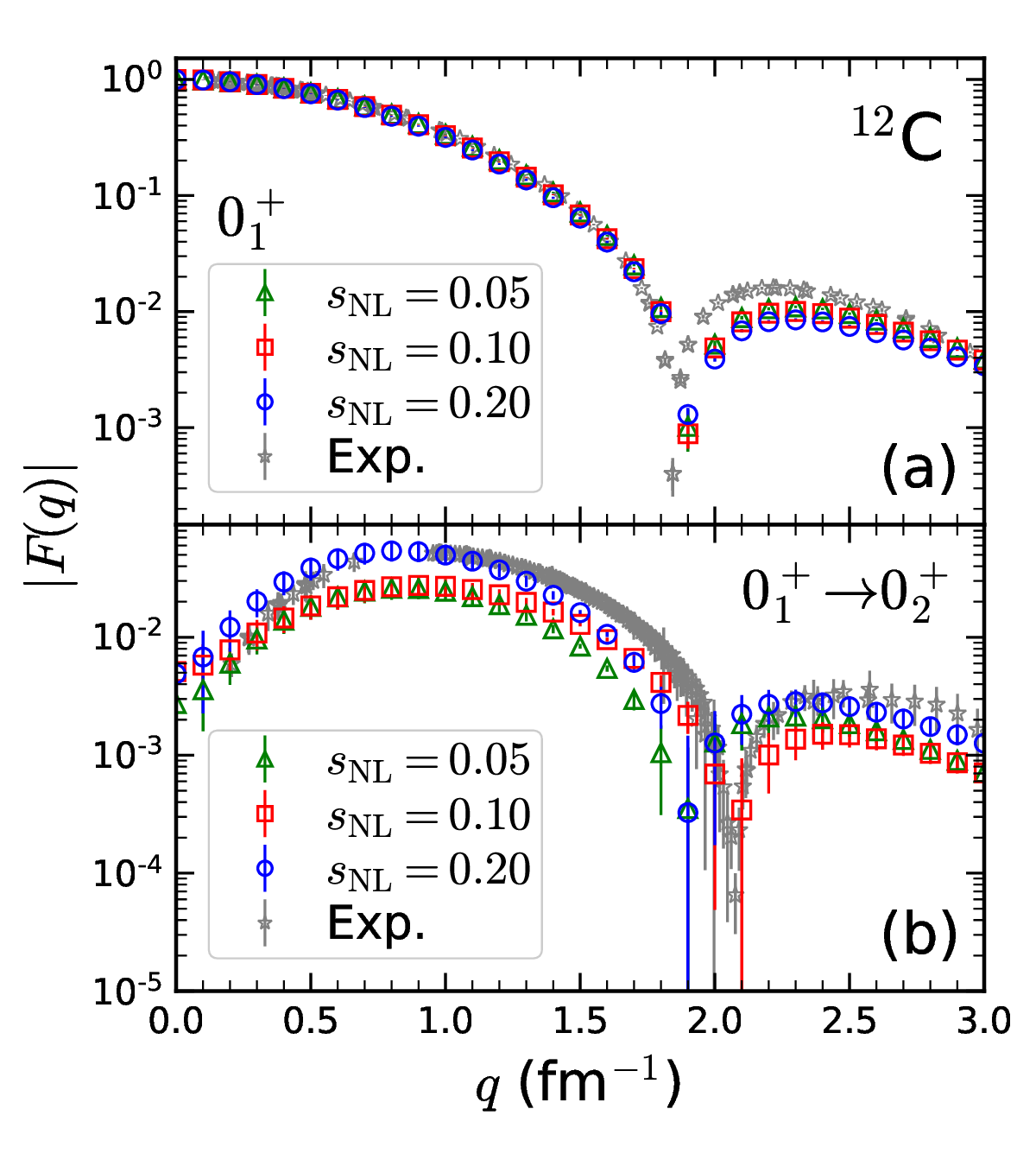}
\caption{\label{fig:form-v3} Absolute value of the form factor of (a) the ground state and (b)
  the transition from the ground state to the Hoyle state calculated at finite Euclidean projection time $t = 0.4$ MeV$^{-1}$ for the three interactions V1, V2 and
  V3 corresponding to $s_{\rm NL} = 0.1, 0.2, 0.05$, respectively, along with experimental data~\cite{Sick:1970ma,Strehl:1970,Crannell:2005wer,Chernykh:2010zu}.}
\end{figure}
For the bound states ($0_1^+$ and $2_1^+$) and those near threshold ($0_2^+, 2_2^+, 3_1^-$), adding the 3NF has little influence. For the higher-lying states, the 3NF in general contributes more repulsion.
Overall, the agreement with experiment is quite good.

In Fig.~\ref{fig:form-v3}, we show the $^{12}$C form factors obtained at finite Euclidean projection time $t = 0.4$ MeV$^{-1}$ for the three interactions V1, V2, and V3, characterized by a different strength of the non-local smearing. For the ground state, as the three interactions are constrained to give a similar radius, the results for the form factor also appear similar.
However, the transition form factor from the ground state to the Hoyle state is more affected. In particular, the shapes and positions of the maxima and minima are shifted.

\subsection*{Electromagnetic observables}

Details on the calculation of electromagnetic observables can be found in the literature~\cite{Bohr:1969}. Here, we only give the final results and discuss issues directly relevant to the lattice calculation. The quadrupole moment $Q$ for a given state with angular momentum $I$ and projection $M$ along the $z$-axis is defined as
\begin{equation}
    Q = \langle I,M=I| \hat{Q} | I,M=I\rangle,
\end{equation}
where the quadrupole moment operator is given by
\begin{equation}
    e\hat{Q} = \int \rho_c(\mathbf{r}) r^2 (3\cos^2\theta-1) d\tau,
\end{equation}
with $e$ the electric charge, $\rho_c$ the charge density distribution and $d\tau$ is the integration measure for the whole space. The reduced transition probability $B(E2)$ is defined as
\begin{equation}
    B(E2, I_1\to I_2) = \sum_{\mu, M_2} | \langle I_2M_2| \mathcal{M}(E2,\mu) | I_1M_1 \rangle |^2.
\end{equation}
The electric quadrupole operator for different $\mu$ components is given by
\begin{equation}
    \mathcal{M}(E2,\mu) = \int \rho_c(\mathbf{r}) r^2 Y_{2\mu}(\theta,\varphi) d\tau,
\end{equation}
with $Y_{2\mu}(\theta,\varphi)$ the pertinent spherical harmonics.

In the above expressions, the coordinate $\mathbf{r}$ for each nucleon is defined in the c.m. frame of the nucleus.  For the quadrupole moment and the $B(E2)$ strength, the operator to be evaluated is $r^2$
(or $x^2, \dots, xy, \dots$, depending on the choice of $\mu$), as for the RMS radius. We evaluate expressions such as
\begin{align}
  \langle r_p^2 \rangle &= \int \prod_{i=1}^A d\mathbf{r}_i |\Psi(\mathbf{r}_1,\mathbf{r}_2,\dots,\mathbf{r}_A)|^2 \left[ \sum_{j=1}^Z \hat{\mathbf{r}}_j - \frac{1}{A}\int \prod_{l=1}^A d\mathbf{r}_{l}' |\Psi(\mathbf{r}_1',\mathbf{r}_2',\dots,\mathbf{r}_A')|^2 \sum_{k=1}^A \hat{\mathbf{r}}_k' \right]^2 \nonumber \\
    &= \frac{1}{A} \int \prod_{i=1}^A d\mathbf{r}_i^{} d\mathbf{r}_i' |\Psi(\mathbf{r}_1,\mathbf{r}_2,\dots,\mathbf{r}_A)|^2 \left[ \left( \sum_{j=1}^Z \hat{\mathbf{r}}_j^{} - \sum_{k=1}^A \hat{\mathbf{r}}_k' \right) \right]^2 |\Psi(\mathbf{r}_1',\mathbf{r}_2',\dots,\mathbf{r}_A')|^2 \nonumber \\
    &~~~~ -\frac{Z}{2A^2} \int \prod_{i=1}^A d\mathbf{r}_i^{} d\mathbf{r}_i' |\Psi(\mathbf{r}_1,\mathbf{r}_2,\dots,\mathbf{r}_A)|^2 \left[ \left( \sum_{j=1}^A \hat{\mathbf{r}}_j^{} - \sum_{k=1}^A \hat{\mathbf{r}}_k' \right) \right]^2 |\Psi(\mathbf{r}_1',\mathbf{r}_2',\dots,\mathbf{r}_A')|^2, \label{eq:rp}
\end{align}
where we consider the proton density with $Z$ the number of protons and $A$ the number of nucleons.

For the quadrupole moment and the $B(E2)$ matrix element have a similar form. For example, when $\mu = 0$ the operator is proportional to
\begin{equation}
    Q \sim 3z^2 - r^2,
\end{equation}
which can be treated along the lines of Eq.~(\ref{eq:rp}).

\subsection*{Tomography of lattice states}
In the NLEFT simulation for a given state with good angular momentum, the wave function is projected onto a
given {\em irrep} of the cubic group. For example, the case of $J^\pi = 0^+$ corresponds to the $A_1^+$ {\em irrep} \cite{Johnson:1982yq}, which entails an equal superposition of all possible rotations of the wave function. In other words, for $0^+$ states there is no preference in the angular
distribution and the density distribution should be spherical.
However, one may inquire as to the intrinsic shape of a $^{12}$C nucleus, without such a superposition of all possible spatial rotations. Here, we adopt the following strategy: With the PA we can obtain the superposition of a large number of coordinates $r_i$ ($i = 1, \dots, A$) for $A$ nucleons in all possible rotations. For the ground state, we already have the (model-independent) information from Fig.~\ref{fig:theta12_dstx0} (a) that an equilateral triangular configuration is preferred. Hence, for each configuration $r_i$ we calculate its principal axis and align it in such a way that the shortest axis coincides with the $x$-axis.
In this way, the three alpha clusters will be located on the $y-z$ plane. This can be achieved by calculating the matrix
\begin{equation}
    R_{ij} = \sum_{i=1}^A
    \left(\begin{array}{ccc}
      x_ix_j & x_iy_j & x_iz_j \\
      y_ix_j & y_iy_j & y_iz_j \\
      z_ix_j & z_iy_j & z_iz_j \\
    \end{array}\right)
\end{equation}
and solving the eigenvalue problem
\begin{equation}
  R v = \lambda v.
\end{equation}
The resulting three eigenvalues correspond to the length along the longest, shortest, and intermediate principal axes, respectively. The eigenvectors can be used to rotate the original distribution so that the longest, shortest, and intermediate principal axes coinciding with the coordinate axes. For example, if we want to align the shortest axis with the $x$-axis, the intermediate axis with the $y$-axis, and the longest axis with the $z$-axis, we let $v_{\rm min},
v_{\rm mid},$ and $v_{\rm max}$ be the corresponding eigenvectors, and the rotation matrix can be constructed as
\begin{equation}
    O = ( v_{\rm min}, v_{\rm mid}, v_{\rm max} ).
\end{equation}
Then one can rotate all particles $i = 1, \dots, A$ from the old coordinates to the new ones as
\begin{equation}
    (x, y, z)_{\rm new} = (x, y, z)_{\rm old} \times O.
\end{equation}

For nuclear states that we have
already identified as having an equilateral triangle shape, we rotate the configuration
along the $x$-axis so that one of the alpha clusters is positioned on the $y = 0$ plan.  We then symmetrize with respect to $0^\circ$, $120^\circ$, and $240^\circ$ rotations.  For nuclear states that we have
already identified as having an obtuse isosceles shape, we identify the $z$ axis as the direction with the longest RMS deviation of
the nucleon positions relative to the c.m. We then rotate the density configurations along the $z$ axis so that the alpha cluster with the
smallest $z$ value has a positive $y$ coordinate.

\begin{figure}[ht!]
\centering
\includegraphics[width=\columnwidth]{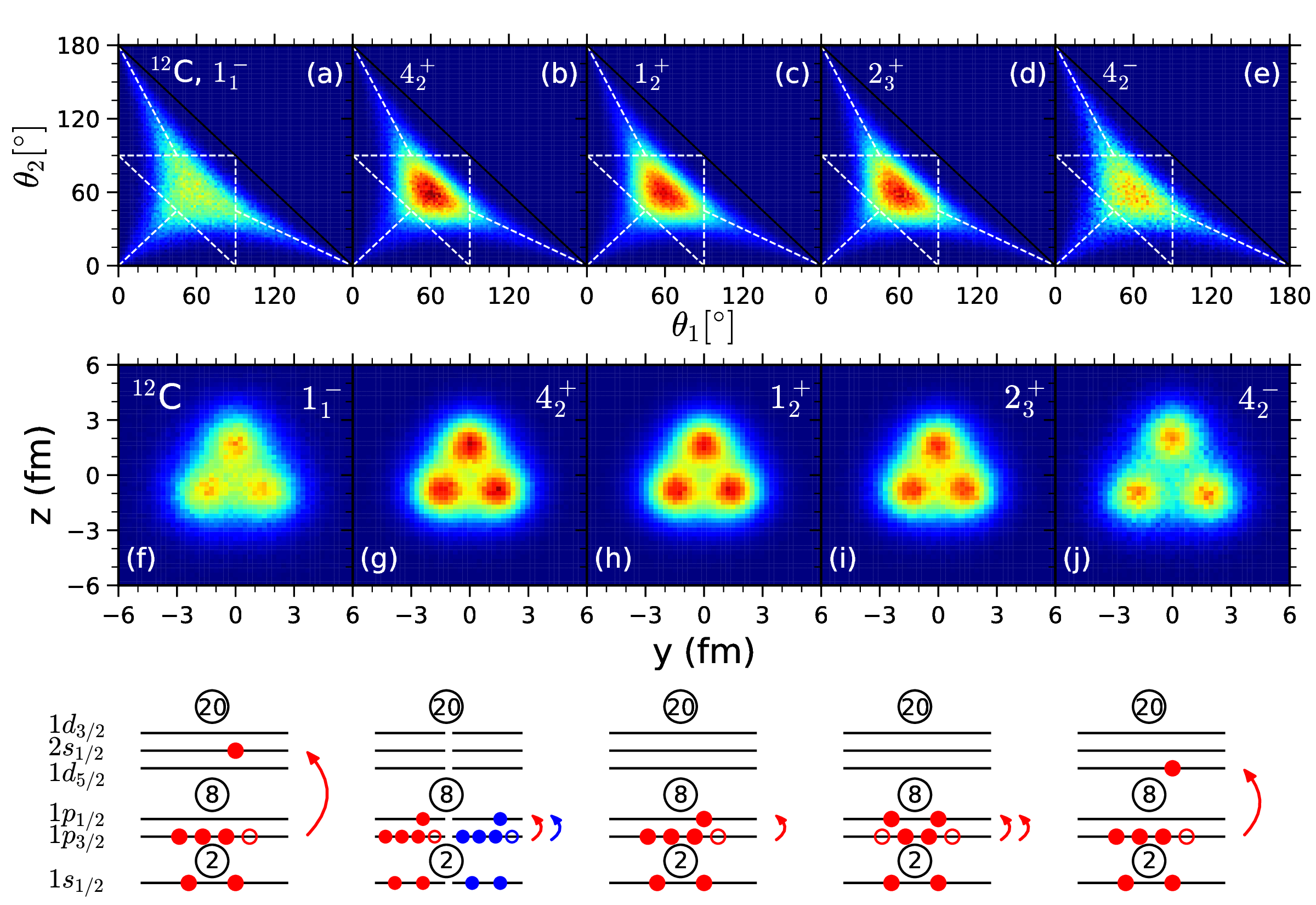}
\caption{\label{fig:dstx0-sm} {\bf Top Panel:} Density distribution for the two inner angles of the triangle formed by the three alpha clusters. {\bf Middle Panel:} Tomographic projection of the nuclear density. {\bf Lower Panel:} Sketch of the orbitals for the shell model initial states used in each of these calculations.}
\end{figure}

In Fig.~\ref{fig:dstx0-sm}, we show the density distribution for the two inner angles of the triangle in the top panel and the tomographic projection of the nuclear density in the middle panel for the states $1^-_1, 4^+_2, 1^+_2, 2^+_3, 4^-_2$ obtained using shell model initial states.  The color scales are same as in Fig.~\ref{fig:theta12_dstx0}.
The corresponding orbitals are shown in the lower panel of Fig.~\ref{fig:dstx0-sm}. The red circles are for protons and blue circles for neutrons, with solid ones for particles and hollow ones for holes.
For those configurations not showing neutrons explicitly, the neutrons are occupied in the lowest possible levels (that is, fully $s_{1/2}$ and $1p_{3/2}$).
They all show a similar equilateral triangle structure as the ground state, but in some cases with a less of a pronounced alpha cluster structure.  It is also interesting to see that among those selected states, the alpha cluster structure is less pronounced when the particle excitation is in $s-d$ shell rather than the $1p_{1/2}$ level.  Furthermore, the separation between the clusters is larger when the particle excitation reaches the $1d_{5/2}$ level.

\begin{figure}[ht!]
\centering
\includegraphics[width=0.49\columnwidth]{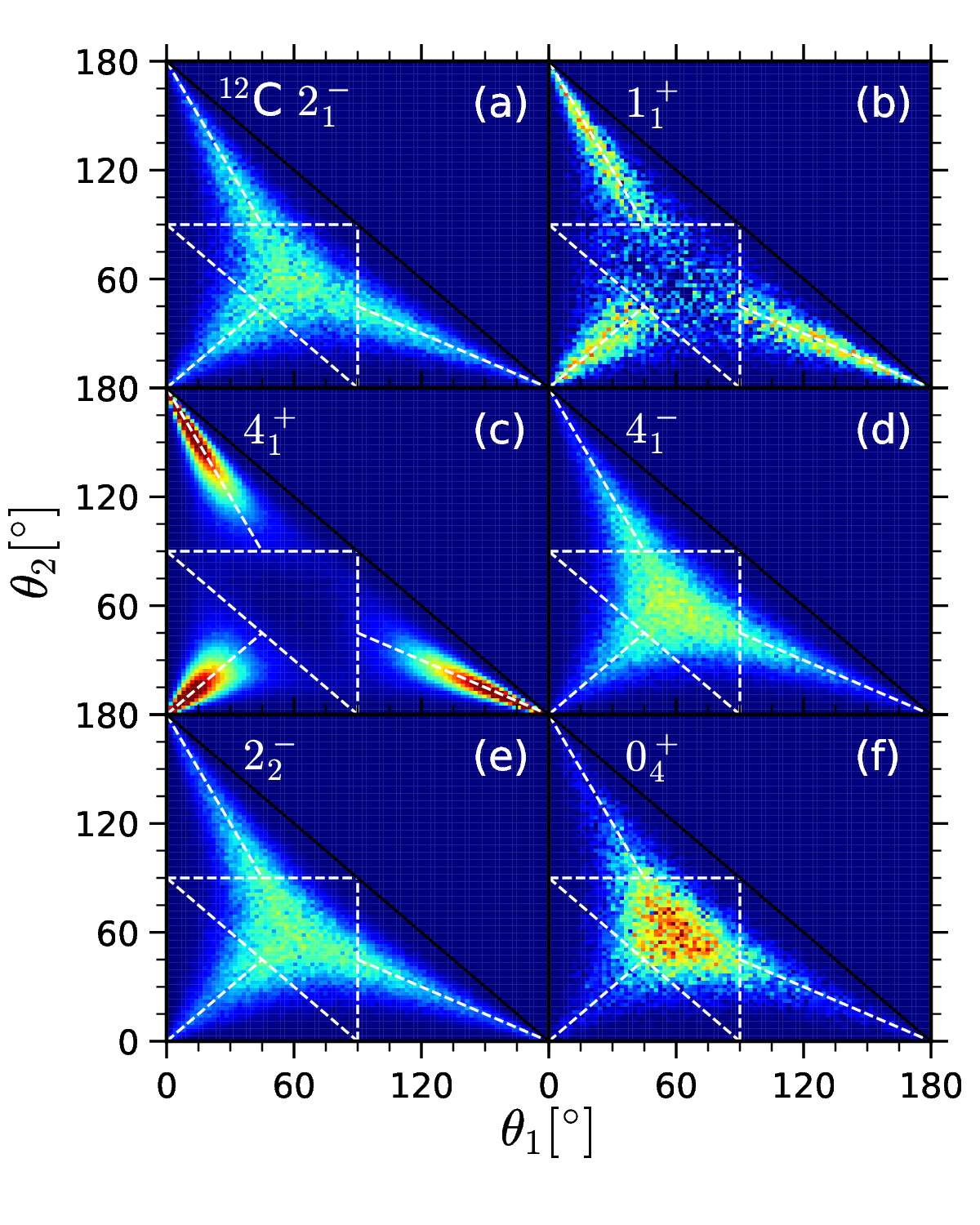}
\includegraphics[width=0.49\columnwidth]{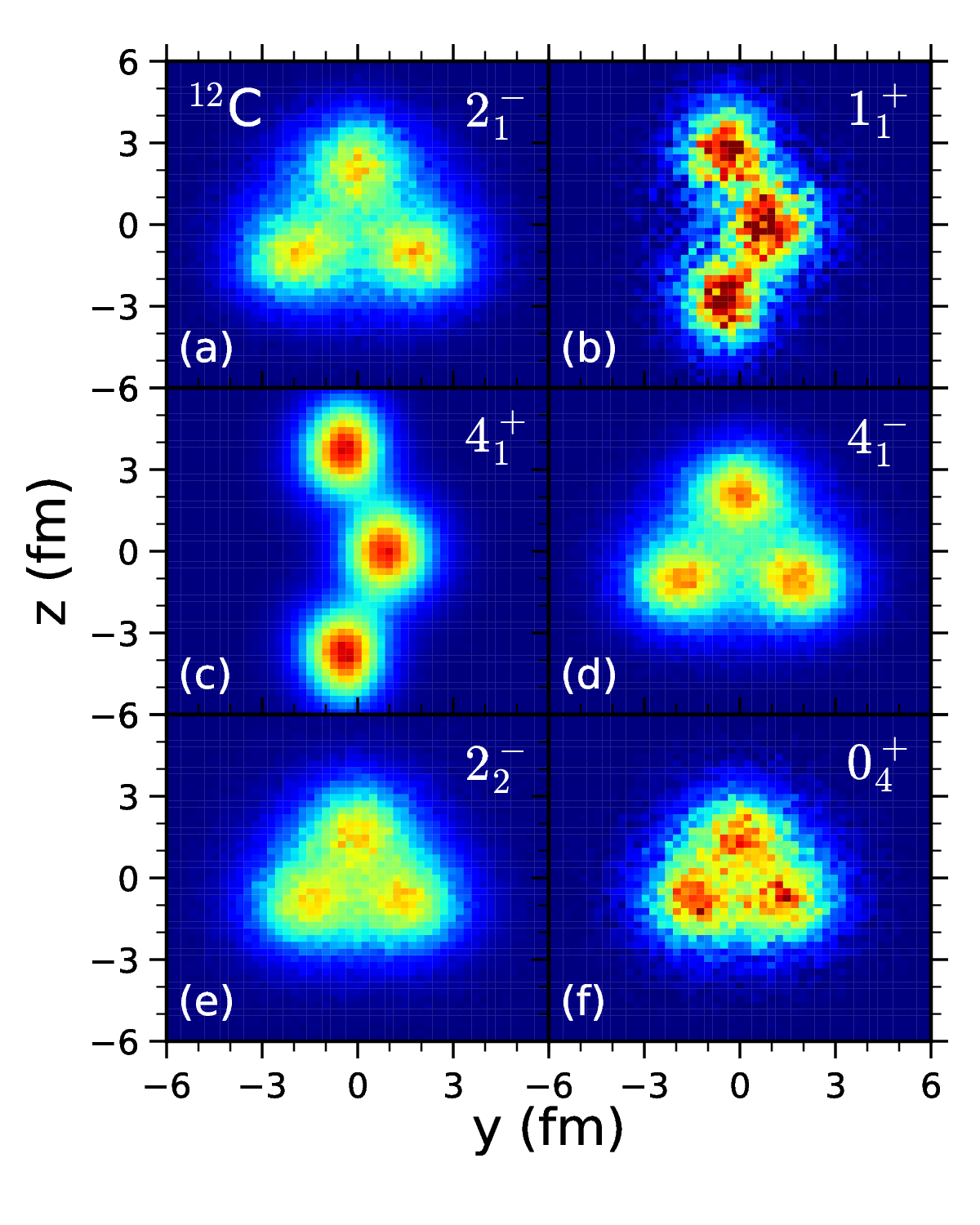}
\caption{\label{fig:dstx0-ext} {\bf Left Panel:} Density distribution for the two inner angles of the triangle formed by the three alpha clusters. {\bf Right Panel:} Tomographic projection of the nuclear density. From (a) to (f), the selected states are ordered by their energies from low to high.}
\end{figure}

For completeness, the density distributions for the other six states calculated in this work are shown in Fig.~\ref{fig:dstx0-ext} and are ordered by excitation energies.
The color scales are the same as in Fig.~\ref{fig:theta12_dstx0}.
The $4_1^+$ appears to be a rotational excitation of the Hoyle state, while the $1_1^+$ is similar to the Hoyle state but with a smaller angle and more compact structure.
The $2_1^-, 1_1^+, 4_1^+$, and $4_1^-$ are obtained with cluster wave functions, while $2_2^-$ is obtained with shell model wave function $2s_{1/2}\otimes 1p_{3/2}$. The $0_4^+$ is obtained using a shell model state with two-particle and two-holes in the $1p_{1/2}\otimes 1p_{3/2}$ orbitals (see the schematic plot in Fig.~\ref{fig:dstx0-sm}).

\begin{figure}[ht!]
\centering
\includegraphics[width=\columnwidth]{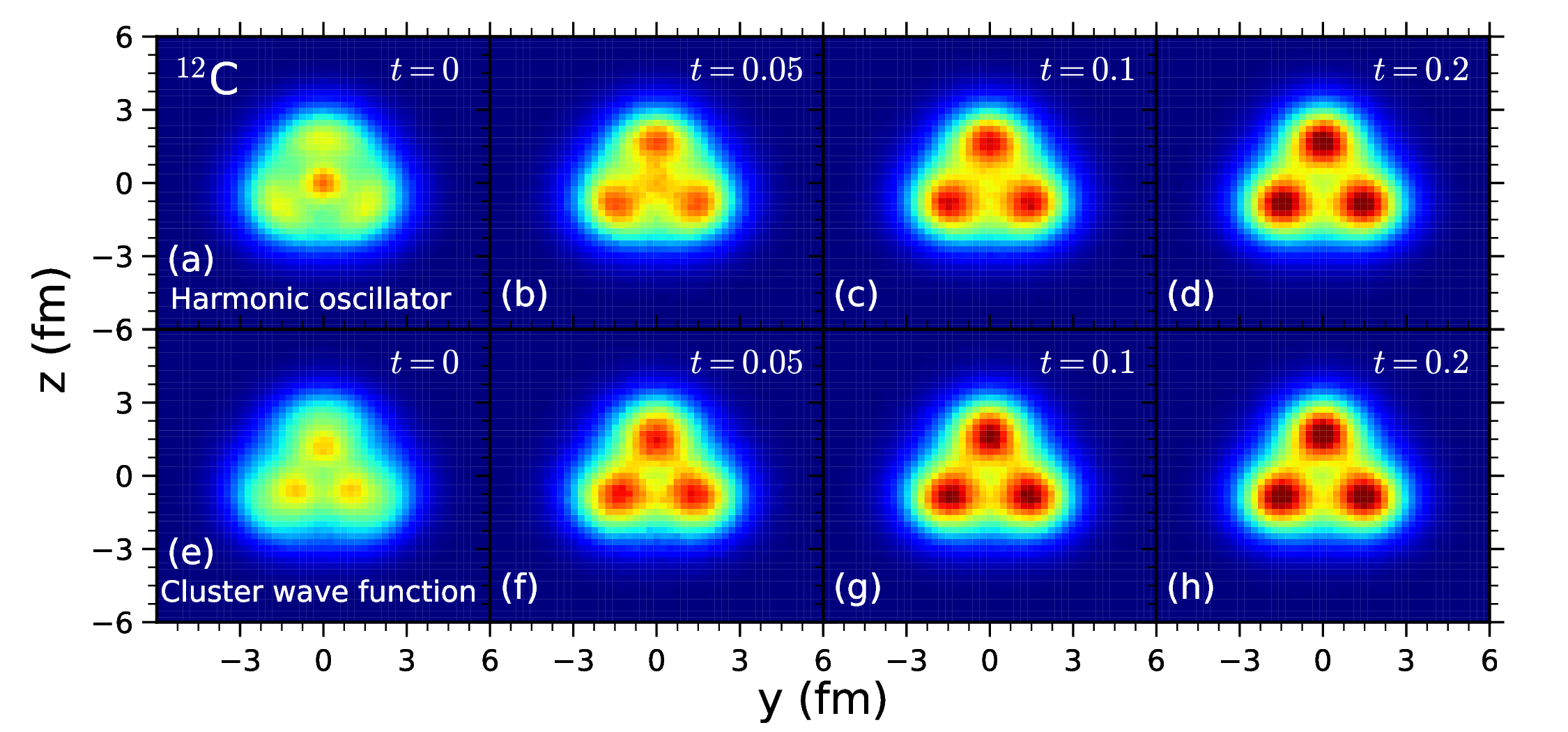}
\caption{\label{fig:dstx0-ho} Tomographic projection of the nuclear density of the ground state of $^{12}$C obtained by NLEFT with Euclidean projection time ranging from $t = 0$ to $0.2$ in units in MeV$^{-1}$. For (a)-(d), the initial state is a shell model wave function composed of harmonic oscillator orbitals. For (e)-(h), the initial state is a cluster wave function.}
\end{figure}

We should emphasize that the intrinsic shapes we are showing are not the result of initial state bias.
This can be seen clearly in Fig.~\ref{fig:dstx0-ho} for the ground state of $^{12}$C.  In the upper
panel (a)-(d), the initial state is a shell model wave function with harmonic oscillator orbitals.
In the lower panel (e)-(h), the initial state is a cluster wave function. We see that the density
distributions from very different initial wave functions begin to look similar as the projection
time increases.  This independence with respect to the choice of initial state was also demonstrated in
previous work~\cite{Epelbaum:2011md,Epelbaum:2012qn,Shen:2021kqr}.  At Euclidean projection time
$t = 0.2\,$MeV$^{-1}$ each of the distributions in the upper and lower panels look similar to the
one shown for the converged ground state in Fig.~\ref{fig:theta12_dstx0}, obtained at larger values of $t$.
At projection time $t = 0$ MeV$^{-1}$, the intrinsic density for the shell model initial wave function
is quite different from the converged ground state. In particular, there is a higher probability
at the center and less probability to be restricted to the $y-z$ plane, and it is therefore more spherical. 

%
%

\subsection*{Model-independent probes of cluster geometry}

\begin{figure}[ht!]
\centering
\includegraphics[width=0.5\columnwidth]{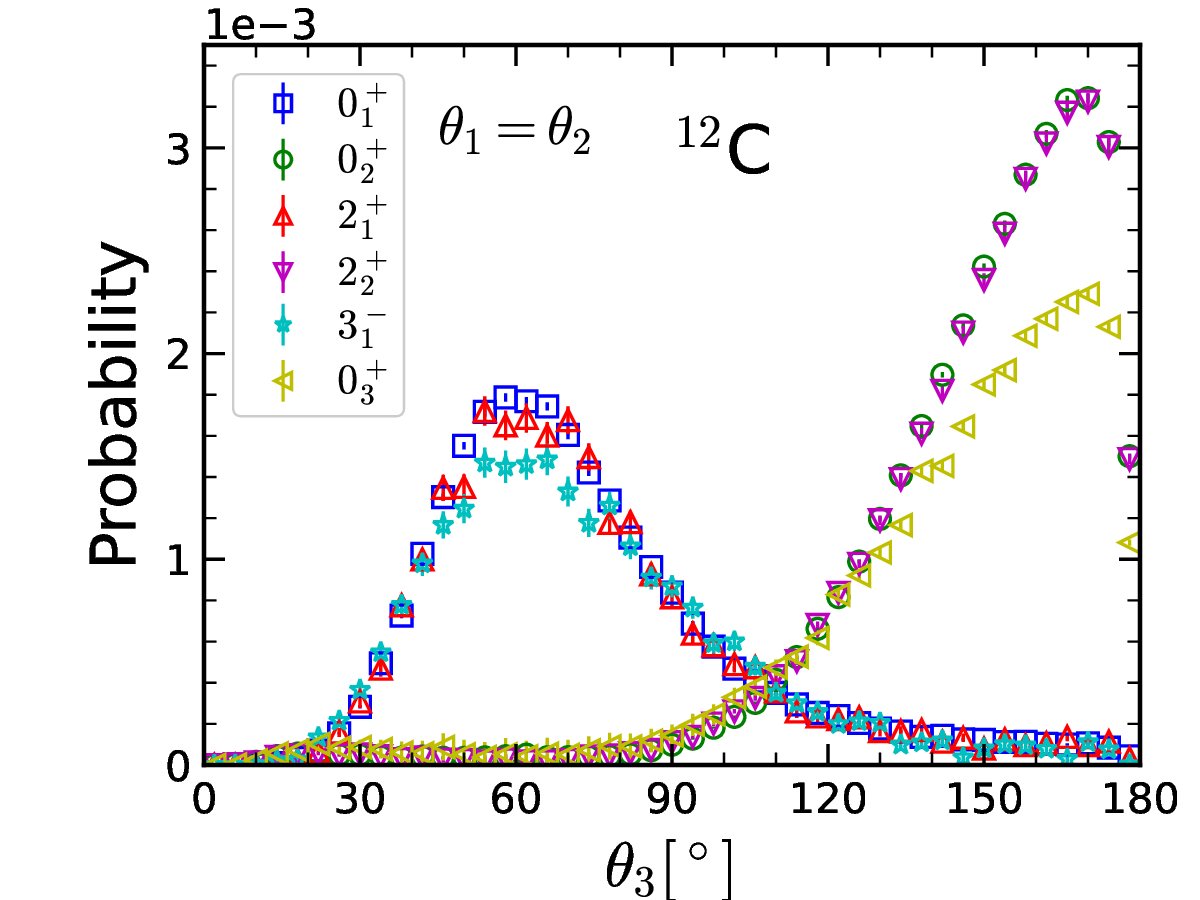}
\caption{\label{fig:theta3} Probability distribution for the third angle of the triangle formed by the
three alpha clusters of $^{12}$C when the other two angles are equal $(\theta_1=\theta_2)$. This coincides with the diagonal line along $\theta_1=\theta_2$ in Fig.~\ref{fig:theta12_dstx0}.}
\end{figure}

To more clearly assess the angular distribution of the various states, we show in Fig.~\ref{fig:theta3} the probability of $\theta_3$
when $\theta_1=\theta_2$, which corresponds to the diagonal line along $\theta_1=\theta_2\in [0^\circ, 90^\circ]$ in Fig.~\ref{fig:theta12_dstx0}. 
For the ground state, we find a clear peak around $60^\circ$ as discussed above, but there is also a non-vanishing probability for an obtuse triangle ($\theta_3>90^\circ$). For the Hoyle state, there is little probability for an acute triangle, with a peak probability around $170^\circ$.
The probability to form an exactly linear chain appears to be strongly suppressed ($\theta_3 \in [176^\circ,180^\circ]$ with
$4^\circ$ resolution). This supports the notion that the intrinsic shape of Hoyle state is a large-angle
obtuse triangle, but not exactly a linear chain. The lowest two $2^+$ states show the same intrinsic shapes as the corresponding $0^+$ states, which gives credit to the assertion made e.g. in Ref.~\cite{Epelbaum:2012qn}
that the $2_2^+$ state is a rotational excitation of the Hoyle state. We also note that the lowest negative-parity state, the $3_1^-$, shows a distribution similar to the ground state.
The angular distribution of the $0_3^+$ state is very similar to the Hoyle state, but with a reduced probability for an exactly isosceles triangles with $\theta_1=\theta_2$. This could be interpreted as a signal of a small-amplitude vibrational excitation. 

%
\begin{figure}[ht!]
\centering
\includegraphics[width=0.5\columnwidth]{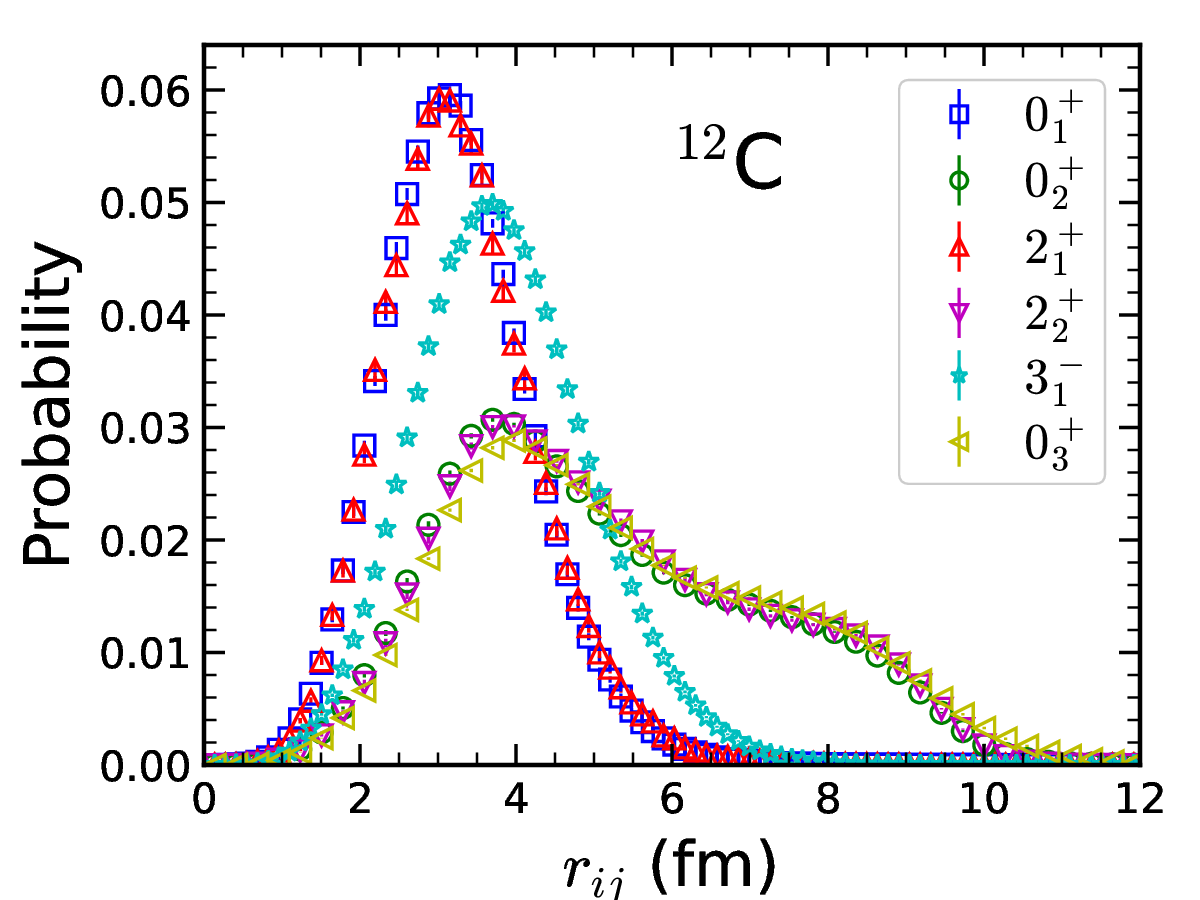}
\caption{\label{fig:dij} Probability distribution for the distance between two alpha particles, for the low-lying states of $^{12}$C.}
\end{figure}
Further model-independent information can be gained from Fig.~\ref{fig:dij}, where the probability distributions for the distance between two alpha particles are shown, for the six lowest states in $^{12}$C. The distance between two alpha particles is state-dependent, for the ground state it peaks at $\sim 3$~fm and
for the Hoyle state near $\sim 4$~fm, as can also seen from the corresponding density distributions in Fig.~\ref{fig:theta12_dstx0}.
For the Hoyle state, one observes an extended tail, which reflects the distance between the two alpha
particles associated with the longest side of the obtuse triangle. Again, we note the stunning similarity in this distribution
for the $0_1^+,2_1^+$ and $0_2^+,2_2^+$ states, reinforcing the notion of the $2^+$ states being
rotational excitations of the corresponding $0^+$ states. Note also that the $3_1^-$ state is somewhat
broader than the ground state, but by far less extended than the Hoyle state.
The similarity of $0_3^+$ and Hoyle state, but with a slightly larger separation distance distribution, again supports the notion of a small amplitude vibrational excitation.

\subsection*{Electromagnetic density distributions}

\begin{figure}[ht!]
\centering
\includegraphics[width=0.5\columnwidth]{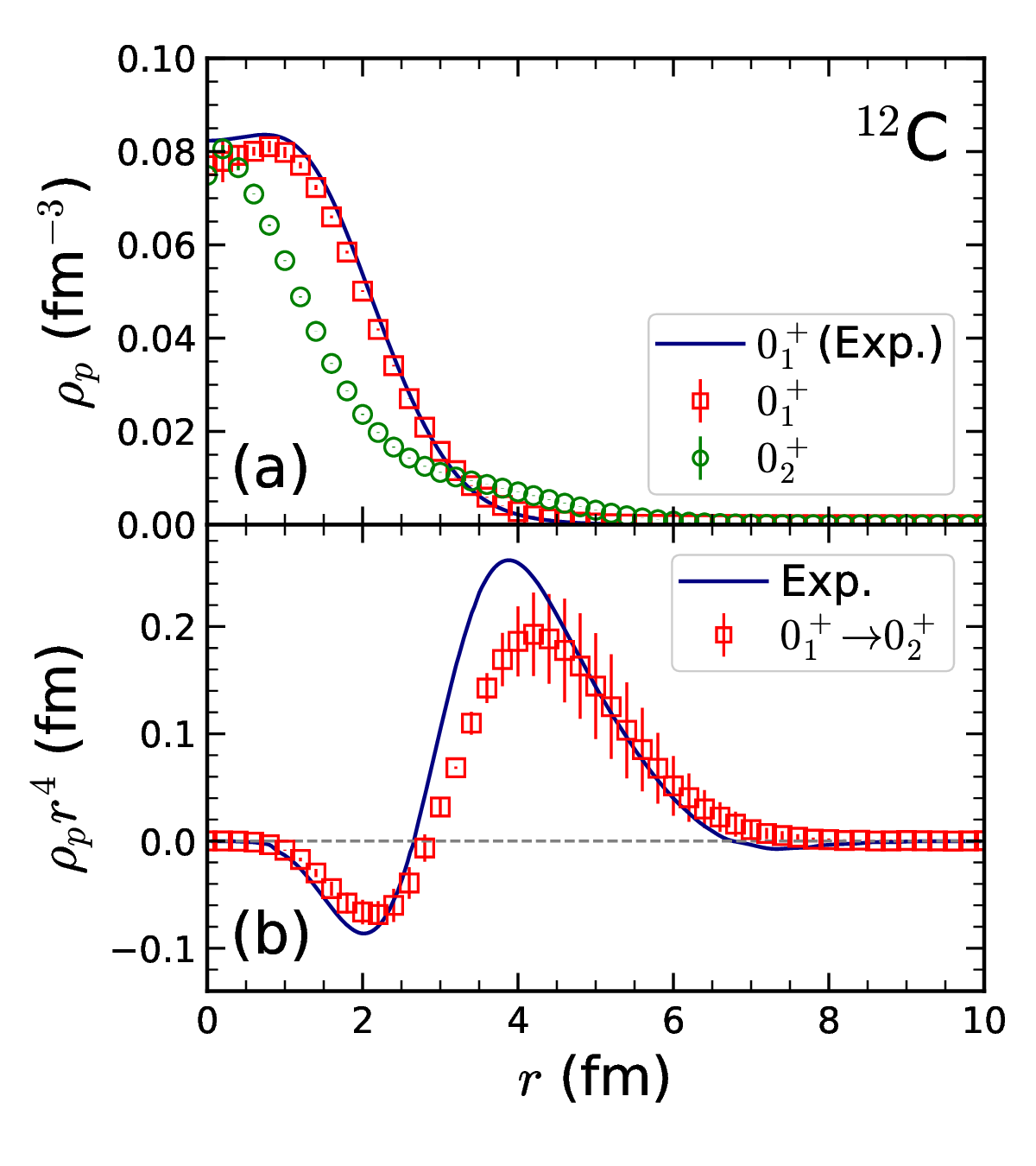}
\caption{\label{fig:rho} Proton densities of $^{12}$C calculated by NLEFT in comparison
with experiment \cite{ncda,Chernykh:2010zu}  for (a) the ground state and the Hoyle state; (b) the transition from the ground state to the Hoyle state.}
\end{figure}
Here, we display the electromagnetic density distribution not shown in the main text.
In Fig.~\ref{fig:rho} the proton radial density distribution of the $0_1^+$, $0_2^+$ states is displayed,
and the transition of $0_1^+\to 0_2^+$ in comparison with the available experimental data~\cite{ncda,Chernykh:2010zu}.
The density distribution of the ground state is nicely reproduced.
The slight decrease in the center at $r = 0$ fm reflects the fact that the three alpha clusters are equally
far away from the center.
For the Hoyle state, the density distribution shown in Fig.~\ref{fig:rho} (a) has no such decrease.
This is also consistent from the shape information, that one alpha cluster is closely located near
the center. The decrease of the density distribution as $r$ increases has a visible slow down from $r = 2\,$fm
to $r = 6\,$fm, and this corresponds to the separation between the two alpha clusters forming the longest side of the obtuse triangle.
%
%
The pattern of the transition density $0_1^+\to 0_2^+$ is also reproduced, with a minimum near $r = 2\,$fm, which is the approximate distance from the center to the $\alpha$ clusters in the ground state.  The maximum occurs near $r = 4\,$fm, which is the approximate distance from the center of the Hoyle state to the two $\alpha$ clusters furthest from the center.



\end{document}